\documentclass[aps,epsfig,graphics,twocolumn,floatfix,mathbbm,a4paper,nofootinbib]{revtex4}

\usepackage{amsmath,amsfonts,amssymb,graphics,graphicx,epsfig,color,times,bbm}

\begin{document}
\bibliographystyle{apsrev}

\newcommand{\R}{\mathbbm{R}}
\newcommand{\rr}{\mathbbm{R}}
\newcommand{\E}{{\cal E}}
\newcommand{\cc}{{\cal{C}}}
\newcommand{\ii}{\mathbbm{1}}

\newcommand{\1}{\mathbbm{1}}
\newcommand{\F}{\mathbbm{F}}

\newcommand{\tr}[1]{{\rm tr}\left[#1\right]}
\newcommand{\gr}[1]{\boldsymbol{#1}}
\newcommand{\be}{\begin{equation}}
\newcommand{\ee}{\end{equation}}
\newcommand{\bea}{\begin{eqnarray}}
\newcommand{\eea}{\end{eqnarray}}
\newcommand{\ket}[1]{|#1\rangle}
\newcommand{\bra}[1]{\langle#1|}
\newcommand{\avr}[1]{\langle#1\rangle}
\newcommand{\D}{{\cal D}}
\newcommand{\eq}[1]{Eq.~(\ref{#1})}
\newcommand{\ineq}[1]{Ineq.~(\ref{#1})}
\newcommand{\sirsection}[1]{\section{\large \sf \textbf{#1}}}
\newcommand{\sirsubsection}[1]{\subsection{\normalsize \sf \textbf{#1}}}
\newcommand{\ack}{\subsection*{\normalsize \sf \textbf{Acknowledgements}}}
\newcommand{\front}[5]{\title{\sf \textbf{\Large #1}}
\author{#2 \vspace*{.4cm}\\
\footnotesize #3}
\date{\footnotesize \sf \begin{quote}
\hspace*{.2cm}#4 \end{quote} #5} \maketitle}
\newcommand{\eg}{\emph{e.g.}~}

\newcommand{\proofend}{\hfill\fbox\\\medskip }

%---------------------------------------------------------------------------

\newtheorem{theorem}{Theorem}
\newtheorem{proposition}{Proposition}

\newtheorem{lemma}{Lemma}

\newtheorem{definition}{Definition}
\newtheorem{corollary}{Corollary}

\newcommand{\proof}[1]{{\bf Proof.} #1 $\proofend$}

\newcommand{\alejo}[1]{{\color{red} #1}}

\newcommand{\uno}{1\!\!1}

\title{Super-activation of quantum nonlocality}

\author{Carlos Palazuelos}
\affiliation{Instituto de Ciencias Matem\'aticas, ICMAT\\Consejo Superior de Investigaciones Cient\'ificas,\\ Campus de Cantoblanco, 28049 Madrid, Spain}

\date{\today}

\begin{abstract}
In this paper we show that quantum nonlocality can be super-activated. That is, one can obtain violations of Bell inequalities by tensorizing a local state with itself. In the second part of this work we study how large these violations can be. In particular, we show the existence of quantum states with very low Bell violation but such that five copies of them give very large violations. In fact, this gap can be made arbitrarily large by increasing the dimension of the states.
\end{abstract}

\maketitle

%%%%%%%%%%%%%%%%%%%%%%%%%%%%%%%%%%%%%%%%%%%%%%%%%%%%%%%%%%%%%%%%%%%%%%

That by combining two quantum objects one can get \emph{something better} than the sum of their individual uses seems to be a characteristic feature of quantum mechanics. In particular, in quantum information theory this effect has been extensively studied in quantum channel theory (see for instance \cite{SmYa}, \cite{Hastings}, \cite{CCH}) and entanglement theory (see for instance \cite{HHH}, \cite{ShSm}). Actually, some of these works show a much stronger behavior called \emph{super-activation}. That is, one can get a quantum effect by combining two objects with no quantum effects. The aim of this work is to study this phenomenon in the context of quantum nonlocality.

The study of quantum nonlocality dates back to the seminal work by Bell (\cite{Bell}). In this work the author took the apparently metaphysical dispute arising from the previous intuition of Einstein, Podolski and Rosen (\cite{EPR}) and formulated it in terms of assumptions which naturally lead to a refutable prediction. Given two spatially separated quantum systems, controlled by Alice and Bob respectively and specified by a bipartite quantum state $\rho$, Bell showed that certain probability distributions $p(a,b|x,y)$ obtained from an experiment in which Alice and Bob perform some measurements $x$ and $y$ in their corresponding systems with possible outputs $a$ and $b$ respectively, cannot be explained by a local hidden variable model (LHVM). Specifically, Bell showed that the assumption of a LHVM implies some inequalities on the set of probability distributions $p(a,b|x,y)$, since then called \emph{Bell inequalities}, which are violated by certain quantum probability distributions produced with an \emph{entangled state}.

Though initially discovered in the context of \emph{foundations of quantum mechanics}, violations of Bell inequalities, commonly known as \emph{quantum nonlocality}, are nowadays a key point in a wide range of branches of quantum information science. In particular, nonlocal probability distributions provide the quantum advantage in the security of quantum cryptography protocols (\cite{AMG}, \cite{ABGMPS}), communication complexity protocols (see the recent review \cite{BCMW}) and in the generation of trusted random numbers (\cite{PCMBM}).

In order to pass from the \emph{probability distribution level} to the \emph{quantum state level}, we say that a bipartite quantum state $\rho$ is \emph{nonlocal} if it can lead to certain quantum probability distributions $p(a,b|x,y)$ in an Alice-Bob scenario violating some Bell inequality. In the case where any probability distribution $p(a,b|x,y)$ produced with the state $\rho$ can be explained by a LHVM, we say that $\rho$ is \emph{local}.

Due to the importance of quantum nonlocality, it is a fundamental problem to study whether the nonlocality of a quantum state can be super-activated. That is,
\begin{align}\label{question}
\text{can the state $\rho\otimes \rho$ be nonlocal if $\rho$ is local?}
\end{align}
Some interesting progress have been made on this problem. Indeed, after some numerical attempts (\cite{LiDo}), two partial answers to question (\ref{question}) have recently been obtained in \cite{CASA} and \cite{NaVe}. In the first work, a positive answer to question (\ref{question}) was given in the multipartite setting and for the restricted case of von Neumann measurements. On the other hand, in \cite{NaVe} a strong super-activation result was given when one is restricted to the particular measurement scenario of two inputs and two outputs per party. Despite this considerable effort, question (\ref{question}) has remained open until now. In this work we show that the general problem (\ref{question}) has a positive answer. Furthermore, as we will explain later, previous results suggest that we can get an \emph{unbounded Bell violation} with the state $\rho\otimes \rho$.

We must mention that some previous results on super-activation have been obtained in different contexts of quantum nonlocality. A remarkable one was given by Peres, who showed that super-activation of two-qubit Werner states can occur when local pre-processing is allowed on several copies of the state of Alice and Bob (\cite{Peres}). Super-activation was also considered for arbitrary entangled states by allowing local pre-processing on the tensor product of different quantum states (\cite{MLD}). In contrast, our results do not make use of any local pre-processing. The problem of super-activation was also studied in the context of tensor networks ([8], [9])."
%However, despite this considerable effort question (\ref{question}) remained open until now.

\section{Probability distributions in a measurement setting}\label{LV(M)}
A standard scenario for studying quantum nonlocality consists of two spatially separated and non-communicating parties, usually called Alice and Bob. Each of them can choose among different observables, labeled by $x=1,\cdots , N$ in the case of Alice and $y=1,\cdots , N$ in the
case of Bob. The possible outcomes of these measurements are labeled by $a=1,\cdots , K$ in the case of Alice and $b=1,\cdots
, K$ in the case of Bob. Following the standard notation, we will refer to the observables $x$ and $y$ as \emph{inputs} and call $a$
and $b$ \emph{outputs}. For fixed $x,y$, we will consider the probability distribution $(P(a,b|x,y))_{a,b=1}^K$ of positive real numbers satisfying $$\sum_{a,b=1}^KP(ab|xy)= 1.$$ The collection $P=\Big(P(a,b|x,y)\Big)_{x,y; a,b=1}^{N,K}$ will be also referred as a \emph{probability distribution}.

Given a probability distribution $P$, we will say that $P$ is \emph{Classical} or \emph{LHVM} if
\begin{align}\label{classical}
P(a,b|x,y)=\int_\Omega P_\omega(a|x)Q_\omega(b|y)d\mathbb{P}(\omega)
\end{align}
for every $x,y,a,b$, where $(\Omega,\Sigma,\mathbb{P})$ is a probability space, $P_\omega(a|x)\ge 0$ for all $a,x,\omega$, $\sum_a
P_\omega(a|x)=1$ for all $x,\omega$ and analogous conditions for the $Q_\omega(b|y)$'s. We denote the set of classical probability distributions by $\mathcal{L}$. On the other hand, we say that $P$ is \emph{Quantum} if there exist two Hilbert spaces $H_1$, $H_2$ such that
\begin{align}\label{quantum}
P(a,b|x,y)=tr(E_x^a\otimes F_y^b \rho)
\end{align}
for every $x,y,a,b$, where  $\rho\in B(H_1\otimes H_2)$ is a density operator and $(E_x^a)_{x,a}\subset B(H_1)$, $(F_y^b)_{y,b}\subset B(H_2)$ are two sets of operators representing positive-operator valued measurements (POVM) on Alice's and Bob's systems. We denote the set of quantum probability distributions by $\mathcal{Q}$.

It is not difficult to see that both $\mathcal{L}$ and $\mathcal{Q}$ are convex sets and, furthermore, that $\mathcal{L}$ is a polytope. The inequalities describing the facets of this set are usually called \emph{Bell inequalities}. As we have explained before, the fact that $\mathcal{L}$ is strictly contained in $\mathcal{Q}$ or, equivalently, that there exist some elements $Q\in \mathcal{Q}$ which \emph{violate certain Bell inequalities}, is a crucial point in quantum information theory. We say that a bipartite quantum state is \emph{local} if for all families of POVMs $\{E_x^a\}_{x,a}$, $\{F_y^b\}_{y,b}$, the corresponding probability distribution $Q=\big(tr(E_x^a\otimes F_y^b \rho)\big)_{x,y;a,b}$ belongs to $\mathcal L$. Otherwise, we say that $\rho$ is nonlocal. It is known that a pure state $|\varphi\rangle\langle\varphi|$ is nonlocal if and only if it is entangled (\cite{GiPe}). However, the situation is not as nice in the case of general states. Indeed, it was shown in \cite{Werner}, \cite{Barret} that there exist certain entangled states $\rho$ which are local, laying the foundation for the later understanding of quantum entanglement and quantum nonlocality as different quantum resources.

In order to separate the sets $\mathcal{L}$ and $\mathcal{Q}$, it is very helpful to slightly extend the notion of Bell inequality. For an arbitrary $M\in \mathbb{R}^{N^2K^2}$, we consider the quotient
\begin{align*}
LV(M)=\frac{\omega^*(M)}{\omega(M)},
 \end{align*}where we define $\omega^*(M)=\sup\{|\langle M,Q\rangle|: Q\in \mathcal Q\}$ and $\omega(M)=\sup\{|\langle M,P\rangle|: P\in \mathcal L\}$ and for every probability distribution $P$ we denote
\begin{align*}
\langle M,P\rangle=\sum_{x,y;a,b=1}^{N,K}M_{x,y}^{a,b}p(a,b|x,y)
\end{align*}(see \cite{JPPVW2}, \cite{JPPVW}, \cite{JP} for a complete study on this). Note that the existence of Bell violations can be stated by: $LV(M)> 1$ for certain $M$'s.

\section{The Khot and Visnoi game}\label{Section: KV game}
In the remarkable paper \cite{BRSW} the authors used a particularly interesting game $G_{KV}$ to give very tight estimates in the context of \emph{large violations of Bell inequalities}. This game is usually called the \emph{Khot-Visnoi game} (or KV game) because it was first defined by Khot and Visnoi to show a large integrality gap for a semidefinite programming relaxation of certain complexity problems (see \cite{KhVi} for details). Since the KV game will play an important role in this work we will give a brief description of it (see \cite{BRSW} for a much more complete explanation). For any $n=2^l$ with $l\in \mathbb{N}$ and every $\eta\in [0,\frac{1}{2}]$ we consider the group $\{0,1\}^n$ and the Hadamard subgroup $H$. Then, we consider the quotient group $G=\{0,1\}^n/H$ which is formed by $\frac{2^n}{n}$ cosets $[x]$ each with $n$ elements. The questions of the games $(x,y)$ are associated to the cosets whereas the answers $a$ and $b$ are indexed by $[n]$. The game works as follows: The referee chooses a coset $[x]$ uniformly at random and one element $z\in \{0,1\}^n$ according to the probability distribution $Pr(z(i)=1)=\eta$, $Pr(z(i)=0)=1-\eta$, independently of $i$. Then, the referee asks question $[x]$ to Alice and question $[x\oplus z]$ to Bob. Alice and Bob must answer with an element of their corresponding cosets and they win the game if and only if $a\oplus b=z$. We can realize the KV game as an element in $\mathbb{R}^{N^2K^2}$ with $N=\frac{2^n}{n}$ and $K=n$. Actually, it is very easy to see that for every probability distribution $P=\big(P(a,b|[x],[y])\big)_{[x],[y]=1;a,b=1}^{N,K}$ we have
\begin{align*}
\langle G_{KV},P\rangle=\mathbb{E}_z\frac{n}{2^n}\sum_{[x]}\sum_{a\in[x]}P\big(a,a\oplus z|[x],[x\oplus z]\big).
\end{align*}Now, as a consequence of a clever use of the hypercontractive inequality, one can see that $\omega(G_{KV})\leq n^{-\frac{\eta}{1-\eta}}$ (see \cite[Theorem 7]{BRSW}). Furthermore, one can define, for any $a\in\{0,1\}^n$, the vector $|u_a\rangle\in \mathbb{C}^n$ by $u_a(i)=\frac{(-1)^{a(i)}}{\sqrt{n}}$ for every $i=1,\cdots, n$. It is trivial from the properties of the Hadamard group that $\big(P_a=|u_a\rangle \langle u_a|\big)_{a\in [x]}$ defines a von Neumann measurement (vNm) for every $[x]$. These measurements will define Alice and Bob's quantum strategies. Then, as was shown in \cite{BRSW}, for $\eta=\frac{1}{2}-\frac{1}{\ln n}$, $Q$ the quantum probability distribution constructed with the maximally entangled state in dimension $n$, $|\psi_n\rangle=\frac{1}{\sqrt{n}}\sum_{i=1}^n|ii\rangle$, and the previous vNms, one obtains
\begin{align}\label{KV-MES}
\omega(G)\leq C\frac{1}{n}   \text{     } \text{     } \text{  and   } \text{     } \text{     }  \langle G_{KV}, Q\rangle\geq C'\frac{1}{(\ln n)^2},
\end{align}where $C$ and $C'$ are universal constants which can be taken to be, respectively, $C=e^{4}$ and $C'=4$ (\cite{footnote}).
%Note that this says, in particular, that $LV_{|\psi_n\rangle}\geq \tilde{C}\frac{n}{(\ln n)^2}$, where $\tilde{C}\geq e^{-3}$.

\section{Super-activation of quantum nonlocality}\label{Section: super-activation}
In order to show our super-activation result let's consider the \emph{isotropic state}
\begin{align}\label{werner state}
\delta_p=p|\psi_d\rangle\langle\psi_d|+(1-p)\frac{\uno}{d^2},
\end{align}where $|\psi_d\rangle\langle\psi_d|$ is the maximally entangled state in dimension $d$ and $\frac{\uno}{d^2}$ is the maximally mixed state. It was proven in \cite{APBTA}, \cite{Barret} that $\delta_p$ is local if $$p=\frac{(3d-1)(d-1)^{(d-1)}}{(d+1)d^d}.$$
Let's fix $d=8$ so that $p=\alpha\frac{1}{d}$ for a certain $\alpha> 1$ and from this point on let us remove the $p$-dependence of $\delta$.

By the previous explanation, it suffices to find a natural number $k$ and a quantum probability distribution $Q$ constructed with the state $\delta^{\otimes_k}$ such that $Q$ does not belong to $\mathcal L$. Therefore, let's consider an arbitrary $k$ and note that $\delta^{\otimes_k}$ can be expanded as
\begin{align}\label{rest of the terms}
p^k|\psi_d\rangle\langle\psi_d|^{\otimes_k}+\cdots\cdots = p^k|\psi_{d^k}\rangle\langle\psi_{d^k}|+\cdots\cdots,
\end{align}where the rest of the terms in Equation (\ref{rest of the terms}) are formed by tensor products of $|\psi_d\rangle\langle\psi_d|$'s and $\frac{\uno}{d^2}$'s with certain coefficients which are products of $p$'s and $(1-p)$'s.

In order to find our violation of a Bell inequality, we will construct the quantum probability distribution and the violated Bell inequality at the same time. Indeed, we will consider the KV game for $n=d^k$, $G_{KV}$, and the associated vNms in dimension $n$. Now, on the one hand, we have said in Section (\ref{Section: KV game}) that
\begin{align}\label{KV upper bound}
\omega(G_{KV})\leq C\frac{1}{d^k}.
\end{align}
Therefore, we will finish our proof by showing that for a high enough $k$, the quantum probability distribution $Q$ constructed with our vNms and the state $\delta^{\otimes_k}$ satisfies
\begin{align}\label{violation KV}
\frac{\langle G_{KV}, Q\rangle}{C\frac{1}{d^k}}> 1.
\end{align}To see this, we first note that $\langle G_{KV}, Q_i\rangle\geq 0$ for every $i$, where $Q_i$ is the quantum probability distribution formed by the vNms and the $i^{th}$ term in (\ref{rest of the terms}). Indeed, this trivially follows from the fact that $G_{KV}$ is a game, so it has, in particular, positive coefficients. Therefore, there will be no cancelations and it is enough to show (\ref{violation KV}) for the first term in (\ref{rest of the terms}). Since the state in the first term is the maximally entangled state we know again from Section \ref{Section: KV game} that $\langle G_{KV}, Q_1\rangle$ is greater or equal than $C'\frac{1}{(\ln n)^2}=C'\frac{1}{(k\ln d)^2}$. Therefore, we obtain
\begin{align*}
\frac{\langle G_{KV}, Q\rangle}{C\frac{1}{d^k}}\geq \frac{p^k\langle G_{KV}, Q_1\rangle}{C\frac{1}{d^k}}\geq \frac{C'}{C}\alpha^k\frac{1}{(k\ln d)^2},
\end{align*}which tends to $\infty$ when $k\rightarrow \infty$ since $\alpha> 1$.
The proof now follows trivially.

%Since we are interested in the foundational point of view of our problem we have not sought any kind of optimality in the above result. In fact, we must point out that one could be a little bit sharper in the previous proof. Indeed, to get our result we have argued that, since there are no cancelations between the terms in (\ref{rest of the terms}), it suffices to deal with the first term. However, the reader should note that a direct calculation of the value $\langle G_{KV}, Q_i\rangle\geq 0$ for every $i$ would give a sharper statement in terms of the required $k$ to get our Bell violation.

\section{Quantifying quantum nonlocality and some sharp upper bounds}\label{Section: Quantification}
Beyond their interest from a foundational point of view, \emph{quantifying quantum nonlocality} is very helpful in quantum information theory. Roughly speaking, if violations of Bell inequalities mean that quantum mechanics is more powerful than classical mechanics, the \emph{amount of Bell violation quantifies how much more powerful it is} (see \cite{JPPVW2},\cite{JPPVW} and \cite{JP} for some recent results in this direction). In order to define a measure of quantum nonlocality for a given state $\rho$, let's denote $\mathcal Q_\rho$ the set of all quantum probabilities constructed with the state $\rho$. Then, for a given element $M\in \mathbb{R}^{N^2K^2}$, we will denote
\begin{align}
LV_\rho(M)=\frac{\omega_\rho^*(M)}{\omega(M)},
\end{align}where $\omega_\rho^*(M)=\sup\Big\{\big|\langle M, Q\rangle\big|:Q\in Q_\rho\Big\}$ and $\omega(M)$ is as defined in Section \ref{LV(M)}. Finally, the key object of study is
\begin{align*}
LV_\rho:=\sup_{N,K}\sup_{M\in \mathbb{R}^{N^2K^2}}LV_\rho(M).
\end{align*}
The quantity $LV_\rho$ was introduced in \cite{JP} as a natural measure of \emph{how nonlocal a state $\rho$ is} (see \cite{Palazuelos} for a more complete explanation). Indeed, since nonlocality usually refers to probability distributions, it is natural to quantify the amount of nonlocality of a state $\rho$ by measuring how nonlocal the quantum probability distributions constructed with $\rho$ can be. $LV_\rho$ measures exactly this. In fact, \cite[Proposition 3]{JPPVW} allows us to write $LV_\rho$ in the following alternative way, which emphasizes its connection to nonlocality: $$LV_\rho=\frac{2}{\pi_\rho}-1,$$where $\pi_\rho$ is the infimum over $N,K$ and $P\in \mathcal Q_\rho$ of
\begin{align*}
\sup\Big\{\lambda\in [0,1]: \lambda P+ (1-\lambda)P'\in\mathcal L \text{   }\text{  for some   } P'\in\mathcal L\Big\}.
\end{align*}
Actually, the KV game was considered in \cite{BRSW} to show that $LV_{|\psi_d\rangle}\geq C \frac{d}{(\ln d)^2}$ for certain universal constant $C$, providing in this way a tight lower bound which almost matches the known upper bound estimate $LV_\rho\leq d$ for any $d$-dimensional state $\rho$ (\cite{JPPVW2}, \cite{Palazuelos}). Furthermore, it was recently shown that we cannot completely remove the $\ln$ factor in the estimate given by Buhrman et al. Specifically, the following result was proven in \cite{Palazuelos}.
\begin{align}\label{sharp upper bound}
LV_{|\psi_d\rangle}\leq D \frac{d}{\sqrt{\ln d}},
\end{align}where $D$ is a universal constant.
As we will show in the following section, beyond their own interest, these logarithmic-like estimates are very useful to obtain results about non-multiplicativity.

\section{Unbounded almost-activation }
According to the previous section, the problem of the multiplicativity of quantum nonlocality could be written as
\begin{align}\label{question2}
\text{is   }\text{      }\frac{LV_{\rho^{\otimes_k}}}{(LV_{\rho})^k}>1  \text{   }\text{   for certain states   } \rho?
\end{align}Here, $k$ is any natural number bigger than 1. The proof presented in Section \ref{Section: super-activation} shows that question (\ref{question2}) is affirmative even for $k=2$ and a state $\rho$ verifying $LV_{\rho}=1$. But, how large can the quotient in (\ref{question2}) be? In this section we will show that if we forget about super-activation and we focus on the multiplicativity properties of the measure $LV_\rho$, we can give a much stronger result than the previous one in terms of the amount of violation. Actually, we will show the following result:

For every $\epsilon> 0$ and $\delta> 0$ we have a state $\rho$ (of a sufficiently high dimension $d$) verifying that
\begin{align}\label{almost activation}
LV_{\rho}< 1+\epsilon \text{   }\text{and}\text{   }LV_{\rho^{\otimes_5}}> \delta.
\end{align}
Note that in this case we can make the quotient in (\ref{question2}) arbitrarily large for a fixed number $k=5$ by considering a state $\rho$ of a sufficiently high dimension. This is very different from the estimate obtained in Section \ref{Section: super-activation}, where the increasing $k$ is necessary to get a large violation. The prize to pay now is that we don't know that our initial state is local, but just \emph{almost local} in terms of Bell violations.

In order to prove this result, let's consider $p=\frac{(\ln d)^{\frac{1}{2}-\alpha}}{d}$, where $\alpha$ is an arbitrary constant in $(0,\frac{1}{2})$ and
\begin{align*}
\xi=p|\psi_d\rangle\langle\psi_d|+ (1-p)\frac{\uno}{d^2}.
\end{align*}Using Equation (\ref{sharp upper bound}) and the fact that the state $\frac{\uno}{d^2}$ is separable we deduce that
\begin{align}
LV_\xi\leq Dp\frac{d}{(\ln d)^\frac{1}{2}}+(1-p)\leq D(\ln d)^{-\alpha}+1.
\end{align}
On the other hand, by the same computations as in Section \ref{Section: super-activation}, we can deduce that, if $Q$ is the quantum probability distribution constructed with the vNms associated to the KV game (see Section \ref{Section: KV game}) in dimension $d^5$ and the state $\xi^{\otimes_5}$, we have that
\begin{align*}
LV_{\xi^{\otimes_5}}\geq \frac{\langle G_{KV},Q\rangle}{\omega(G_{KV})}\geq p^5\frac{C'}{C}\frac{d^5}{(5\ln d)^2}=C''(\ln d)^{\frac{1}{2}-5\alpha}.
\end{align*}Taking $\alpha=\frac{1}{11}$ the statement follows by considering a high enough $d$.

\section{Conclusions}
In this work we have proven that quantum nonlocality can be super-activated. This answers a fundamental question about one of the most puzzling and powerful effects in nature. In particular, we have answered the recent enhancement of problem $21$ posed by Liang in the Hannover List of Open Problems in Quantum Information (\cite{Hannover}). Actually, the proof we have presented in this work is very simple and, hopefully, completely understandable for a general audience.
%As we have pointed out in Section \ref{Section: super-activation}, some sharper calculations in the proof we have presented would lead us to a more precise statement.

Beyond the proof of this fundamental result, one could ask about the amount of Bell violation in this super-activation effect. We have shown that the amount of Bell violation attainable by a quantum state is a highly non-multiplicative measure. Note that the enhancement of a Bell violation
via tensor products had already been studied in \cite{LiDo}. However, the enhancement known for mixed states was very mild. Here, we have shown that one can get arbitrarily large Bell violations by taking a finite number of tensor products of an almost-local state. Some results support the conjecture that this phenomenon is also true when we study super-activation, so that one could obtain an unbounded super-activation result; this would mean that one can obtain an arbitrarily large amount of Bell violations by taking a finite number of tensor products of a local state. Indeed, Equation (\ref{sharp upper bound}) strongly supports that a logarithmic-like estimate like the one given in \cite[Equation (12)]{APBTA} for von Neumann measurements should hold for general POVMs. The proofs we have presented above could be then followed step by step to show an unbounded super-activation result. However, at the moment of this writing we do not know how to adapt our techniques in \cite{Palazuelos} to get such an estimate.

Finally, it is worth mentioning that, since the quantum probability distributions that we have used in all our proofs are constructed with vNms (the ones used in the KV game), one can obtain unbounded super-activation of quantum nonlocality in the restricted setting of vNms. Indeed, using the estimate $p_L^\phi\geq \Omega(\frac{\ln d}{d})$ obtained in \cite{APBTA} for vNms, we can follow exactly the same steps as in the previous proofs to obtain an arbitrarily large amount of Bell violation with a finite number of tensor products of a state which is \emph{local under vNms}. However, we must mention that restricting to vNms in the study of activation of quantum nonlocality (or, in general, problems involving tensor products of states) distorts the problem quite a lot.

We would like to thank Y.-C. Liang, M. Navascues, T. Vidick and R. de Wolf for many helpful discussions on previous versions of this article. We would like to thank M. L. Almeida for sharing some personal notes on \cite{APBTA}.

Author's research was supported by EU grant QUEVADIS, Spanish projets QUITEMAD, MTM2011-26912 and MINECO: ICMAT Severo Ochoa project SEV-2011-0087 and the ``Juan de la Cierva'' program.


\begin{thebibliography}{99}
\bibitem{footnote} We are assuming $n\geq 8$, but this assumption will be natural in our proof.
\end{thebibliography}
\end{document}